\documentclass[preprint,showpacs,preprintnumbers,amsmath,amssymb,longbibliography]{revtex4-1}
\usepackage[top=2.5cm]{}
\usepackage{graphicx}
\usepackage{mathrsfs,amsmath}
\usepackage{dsfont}                      
\usepackage{chngcntr}
\usepackage{bm}
\newcommand{\D}{{\mathds{D}}}
\newcommand{\C}{\mathds{C}}
\newcommand{\Ro}{\text{Ro}}
\newtheorem{proposition}{Proposition}
\newenvironment{hypothesis}[1]{\newline\textbf{[Hypothesis #1]}}{\newline}
\usepackage{xcolor}
\usepackage{booktabs}
\usepackage{multirow}

\usepackage{graphicx}
\usepackage{placeins}

\begin{document}

\title{Generalization of Taylor's formula to particles of arbitrary inertia}
\author{S. Boi$^{1}$, A. Mazzino$^{2,3,4}$,  P. Muratore-Ginanneschi$^1$ and S. Olivieri$^{2,3}$}

\begin{abstract}

One  of the  cornerstones  of turbulent  dispersion  is the  celebrated
Taylor formula.   This formula expresses the  rate of transport
(i.e. the  eddy diffusivity) of a  tracer as a time
integral  of the  fluid velocity  auto-correlation function  evaluated
along the fluid trajectories.  
Here, we review the hypotheses which permit to extend Taylor's formula 
to particles  of any  inertia. The hypotheses are independent of the details 
of the inertial particle model.
We also show by explicit calculation, that the hypotheses encompass cases 
when memory terms such as Basset  and  the Fax\'en  corrections  are  
taken into  account in the modeling of inertial particle dynamics.  
\end{abstract}


\maketitle
\section{Introduction}

In the early 20s of last century Sir Geoffrey Ingram Taylor derived
what can be fairly considered one of the cornerstones of large-scale 
transport of tracer particles in fluid  flows \cite{Tay22}. Tracer particles 
are small particles  not affecting the advecting velocity 
field with their  motion:
{\color{black}
\begin{equation}
\frac{d\boldsymbol x}{dt}=\boldsymbol u(\boldsymbol x,t)
\end{equation}
In the the limit  of long observation  time and in suitable conditions, Taylor observed that 
the mean square  of tracer particles displacement behaved  linearly in  time with a  
coefficient now usually referred to as the eddy-diffusivity coefficient ( see e.g. 
\cite{BiCrVeVu95,Frisch,Maz1997,MaMuVu05,CeMaMuVu06,BoMaLa16} ):
\begin{equation}
\left\langle  || \boldsymbol x(t)-\langle\boldsymbol x(t)\rangle||^2\right\rangle\sim 2D t
\end{equation}
Based on this observation, Taylor proceeded to establish a first principle identity 
expressing the tracer particle eddy diffusivity as  a time integral of
the fluid velocity   auto-correlation  function   evaluated   along  the   fluid
trajectories:
\begin{equation}
\label{OldTaylor}
D=\lim_{t\to\infty}\int_0^{t} ds \left\langle \delta   \boldsymbol u(\boldsymbol x(t),t)\cdot\delta\boldsymbol u(\boldsymbol x(s),s)\right\rangle
\end{equation}
where $ \delta\boldsymbol u(\boldsymbol x(t),t)=\boldsymbol u(\boldsymbol x(t),t)-\langle\boldsymbol u(\boldsymbol x(t),t)\rangle$.

Since then, the relation (\ref{OldTaylor}) now going under the name of Taylor's formula has played a key role in the analysis of turbulent dispersion  of tracers \cite{Kra87,MK99}.}
We refer e.g. to 
chapter~\textbf{12} of the textbook \cite{KuCoDo12}
for a review including an introduction to the vast existing literature.


Tracer dispersion  is a small sub-set of a much larger class of transport 
problems: the transport of inertial particles. Inertial particles are small particles
having a finite size and/or different density from that of the carrier fluid 
where they are suspended \cite{ToBo09}.  
Inertial   particles  are   encountered   practically
everywhere, from  our atmosphere (e.g., affecting  the Earth's climate
system because of its effect  on global radiative budget by scattering
and  absorbing  long-wave  and short-wave  radiation  \cite{IPCC};  or
leading to  increased droplet collisions  and the formation  of larger
droplets with a  key role for rain initiation  \cite{FFS01,Mazz1,Mazz2}) and ocean
(e.g.  in  relation  to  phytoplankton  dynamics  in  turbulent  ocean
\cite{FL2015})  to  astrophysics  (in  relation  to  planet  formation
e.g. \cite{BrChPrSp99,WiMeUs08}).

It is therefore not surprising that deriving extensions of Taylor's formula to 
inertial particle dynamics has stirred interest for now already half a century 
\cite{Tch47,Cha64}. A review of early results can be found in \cite{Hin75}. 
Furthermore, most of the existing analytic investigations of inertial particles 
(e.g. \cite{Ree77,PiNi78,Ree88,MeAdHa91} see also \cite{WaSt93} for further references)
use Taylor's formula as a key ingredient.  These works often set out to derive methods for iteratively 
solving coupled systems equations governing the  fluid Eulerian and Lagrangian correlation functions. 

Our aim here is to review in a model-detail independent fashion the conditions presiding over the expression of 
the inertial particle eddy diffusivity as an integral of the  correlation  functions of fluid velocity and external forces  evaluated   along  the   fluid
trajectories.  

There are two closely intertwined reasons why we think that this is interesting.
First, the last decades have seen major developments in the experimental techniques 
to measure  {\color{black} Eulerian fluid flows under real conditions. The best example are the
  sea surface currents which can be determined (as a spatio-temporal  field) via high-frequency radars (see e.g.\cite{Corgnati}). Once a detailed Eulerian fluid flow field is known,
  the determination of the eddy diffusivity via a generalized Taylor formula holding for inertial particles seems to be a very  powerful tool. The reason is that from the space structure of the Eulerian fluid flow 
  one can heuristically argue the properties of the large-scale transport (i.e of the eddy diffusivity). By way of example, a fluid flow having closed structures (i.e. rolls) is expected
to trap inertial particles thus causing a reduction of the transport with respect to flows with open streamlines.  Such kinds of arguments have been successful in the tracer case to identify the so-called constructive and destructive interference regimes \cite{Interf}. The same way of reasoning could now be applied to inertial particles once a generalized Taylor formula is made available.

Similar arguments can be used also in the presence of external forces, which are functions depending - even nonlinearly - on the flow field itself or the particle trajectories explicitly. This brings us to the second reason of this work.} The exact form and relative importance of the forces 
exerted on inertial particles has indeed been object of controversy since the work \cite{Tch47}. 
In more recent years a consensus seems to have been reached based on the
first principle analysis of \cite{MaRi83} and the inclusion of the correction
term advocated in \cite{Auton,AuHuPr88}. {\color{black} A possible review of the evolution history of such models is available in \cite{Michaelides}.} Nevertheless, a model-detail independent analysis 
is justified as it provides a framework to {\color{black}assess}
 the relative importance for diffusion 
of the correction terms distinguishing models of inertial particle dynamics. {\color{black} Thanks to our generalized Taylor formula, one can evaluate the auto-correlations and the cross-correlations of flow and external forces through available data. This allows investigating how and in what regions of the flow the several terms and their mutual interactions contribute to transport,thus providing more physical information about the problem. Moreover, whenever an analytical calculation of the trajectories is  available, it becomes possible to compute exactly the variation  of the eddy diffusivity caused by external forces  and correction terms of the dynamical model. By way of example,  we will consider  the effect of Coriolis, Lorentz, Fax\'en, and lift forces, and in some simple cases we will see how these forces can increase or decrease asymptotic transport, even hindering the molecular diffusion. }


The paper is organized as follows: in Section~\ref{sec:GT} we analyze the hypotheses leading to
the derivation of {\color{black} generalized} Taylor's formula for a wide class of models of inertial particle dynamics.
Technical aspects of this analysis are deferred to an appendix. An important advantage of a model-detail independent 
derivation is to ease the inclusion of the effect of external forces in {\color{black} generalized} Taylor’s formula. 
We avail ourselves of this fact to analyze specific models of inertial particle transport.
In Section~\ref{sec:BBO} we apply the general result to the Basset-Boussinesq-Oseen  model for
inertial particles, with several dynamic scenarios which can be useful
for  applications. In  Section~\ref{sec:MR} we derive Taylor's formula for the 
Maxey-Riley model. This is a refinement of the expression used in \cite{MeAdHa91} 
which retained only leading orders in the expansion in powers of the Stokes number. Finally,
conclusions are presented in Section~\ref{sec:end}.

\section{Generalized Taylor's formula for inertial particle transport}
\label{sec:GT}

We consider a general model of mutually non-interacting inertial particles in a carrier flow.
The state of a single inertial particle is specified by its position $\boldsymbol{\xi}(t)$ and velocity $\boldsymbol  v(t)$ 
at time $t$.  We denote by  $  \boldsymbol u$ the carrier flow, 
a vector field joint function of space and of time variables.
We suppose that the dynamics is amenable to the form of a system of integro-differential 
equations in $d$-spatial dimensions
{\color{black}
\begin{subequations}
\label{GT:eq}
\begin{eqnarray}
\label{GT:v}
&&\boldsymbol{\dot{\xi}}(t)=\boldsymbol{v}(t)
\\
&&\boldsymbol v(t)= \boldsymbol\sigma(\boldsymbol\xi(0),\boldsymbol{v}(0),t)+ \int_0^t\mathrm{d}s\,  \mathds{K}_0 (t-s) \, \boldsymbol  u(\boldsymbol
\xi(s),s)  +   \sum^N_{i=1}\int_0^t\mathrm{d}s  \,  \mathds{K}_i(t-s) \,   \boldsymbol
f_i(\boldsymbol \xi(s),s) 
\label{prima}
\end{eqnarray}
\end{subequations}
As shown in the next sections, many of the current models in literature for the displacement dynamics can be couched into the form (\ref{prima}), whereas this does not happen for models describing the angular dynamics (see e.g. \cite{Candelier}). In (\ref{prima}), we will suppose the transient contributions $\boldsymbol\sigma(\boldsymbol\xi(0),\boldsymbol{v}(0),t)$ depending on the initial condition do not play any role in the asymptotic diffusion and will thus be ignored in the following. The models herein considered fulfill this assumption. Also, the memory of the initial conditions is supposed to be lost in the diffusion dynamics. This  holds true if the  particle-velocity correlation function between two times $t_1$ and $t_2$ is stationary at least asymptotically. That is, it must only depend  on their difference $|t_2-t_1|$ at least when $t_1$ and $t_2$ are sufficiently large. This fact will be crucial in the hypotheses we will be stating later on.}  The vectors $\boldsymbol{f}_i$ $i=1,\dots,N$ stand for 
external forces {\color{black} per unity of mass} acting on the particle such as the  buoyancy,  the Brownian, and the  Coriolis forces.
The detailed form of the $d\times d$-real-matrix-valued integral kernels  $\mathds{K}_0$ and  $\mathds{K}_i$ 
is not important for the analysis of the current section. Drawing on \cite{Cha64} 
we, however, require that
\begin{hypothesis}{I}
\label{hyp:1}
\emph{the integral kernels are \textbf{stationary} and have \textbf{absolutely integrable} components}
\begin{eqnarray}
\label{KappaStar}
\int_{0}^{\infty}\mathrm{d}t\,|\mathds{K}_{i}^{m\,n}(t)|<K_{\star}< \infty\hspace{1.0cm}\forall\,m,n=1,\dots,d\hspace{0.2cm}\&\hspace{0.2cm}\forall i=0,\dots,N
\label{GT:integrable}
\end{eqnarray}
\end{hypothesis}
The hypothesis implies the existence of the Fourier--Laplace transforms
\begin{eqnarray}
\mathds{\hat{K}}_{i}(z)=\int_{0}^{\infty}\mathrm{d}t\,e^{-z\,t}\,\mathds{K}_{i}(t)\,,\hspace{1.0cm}\operatorname{Re}z>0\,,
\hspace{0.5cm}i=0,\dots,N
\nonumber
\end{eqnarray}
Our aim is to compute the inertial particle eddy-diffusivity tensor {\color{black}, which is well-defined and related to asymptotic diffusion whenever the velocity correlation function is stationary at least asymptotically: }
\begin{equation}
\label{eddy1}    
\D=\lim_{t\uparrow\infty}\frac{1}{2\,t}
\Big{\langle}     
\big{(}\boldsymbol{\xi}(t)-\langle\boldsymbol{\xi}(t)\rangle \big{)}   
\otimes      
\big{(}\boldsymbol{\xi}(t)-\langle\boldsymbol{\xi}(t)\rangle \big{)}
\Big{\rangle} 
\end{equation} 
In (\ref{eddy1}) the symbol $\otimes$ denotes the tensor product of vectors and $\langle\dots\rangle$ 
stands for an ``ensemble average''. Ensemble average means here average over any source of 
randomness in the model (e.g. initial data, parameter uncertainty or random carrier velocity field).
By (\ref{GT:v}) we can always couch the eddy diffusivity into the equivalent form
\begin{eqnarray}
\label{eddy2}       \D
=\lim_{t\uparrow\infty}     \operatorname{Sym}  \int_0^t      \mathrm{d}s\,     
\langle\,\delta\boldsymbol{v}(t)\otimes\delta\boldsymbol{v}(s)\rangle
\end{eqnarray}
where
\begin{eqnarray}
\delta\boldsymbol{v}(t)\equiv \boldsymbol{v}(t)-\langle\boldsymbol{v}(t)\rangle
\nonumber
\end{eqnarray}
and $\operatorname{Sym}$ stands for the tensor symmetrization operation
\begin{eqnarray}
\operatorname{Sym}      
\langle\,\delta\boldsymbol{v}(t)\otimes\delta\boldsymbol{v}(s)\rangle=
\frac{\langle\,\delta\boldsymbol{v}(t)\otimes\delta\boldsymbol{v}(s)
+
\delta\boldsymbol{v}(s)\otimes\delta\boldsymbol{v}(t)
\rangle}{2}
\nonumber
\end{eqnarray}
The qualitative reason why the eddy diffusivity is an important indicator of particle 
motion is given by the central limit theorem \cite{Tay22}. 
If the particle  velocity autocorrelation  function decays
sufficiently   fast, one expects that $\boldsymbol{\xi}(t)$ {\color{black} becomes approximately Gaussian for   large 
  times, and the variance is specified by
the  eddy diffusivity tensor (\ref{eddy1}).  }
It should be, however, recalled that the existence of a finite limit for 
(\ref{eddy1}) is not always granted. There are 
physical systems for which $\D$ may vanish (sub-diffusion) or diverge (super-diffusion) 
see e.g. \cite{CaMaMGVu99}.

Here we do not assume directly the existence of (\ref{eddy1}) but we aim to derive it
as a consequence of hypotheses made at the level of the second order statistics of the carrier velocity 
field and the external forces evaluated along particle trajectories.
 
We start by defining the set of Lagrangian $d\times\,d$-matrix-valued correlation functions
\begin{eqnarray}
\tilde{\C}_{i j}(t,t^{\prime})=
\Big{\langle} \boldsymbol \phi_i(\boldsymbol \xi(t),t)\boldsymbol 
\otimes 
\boldsymbol\phi_j(\boldsymbol \xi(t^{\prime}),t^{\prime})
\Big{\rangle}
\hspace{1.0cm} i,j=0,\dots\,N
\label{GT:nz}
\end{eqnarray}
where
\begin{equation}
\boldsymbol\phi_i(\boldsymbol \xi(t),t)=\left\{
\begin{aligned}
\boldsymbol u(\boldsymbol \xi(t),t) -\Big{\langle} \boldsymbol u(\boldsymbol \xi(t),t)\Big{\rangle}\quad &\text{if }i=0&\\
\boldsymbol f_i(\boldsymbol \xi(t),t)-\Big{\langle} \boldsymbol f_i (\boldsymbol \xi(t),t)\Big{\rangle}\quad &\text{if }i=1,\dots,N&\\
\end{aligned}\right.
\end{equation}
Upon recalling (\ref{prima}), it is straightforward to verify that
\begin{eqnarray}
\lefteqn{
\operatorname{Sym} \int_0^t      \mathrm{d}s\,     
\langle\,\delta\boldsymbol{v}(t)\otimes\delta\boldsymbol{v}(s)\rangle
=}
\nonumber\\
&&\sum_{i j=0}^{N}\operatorname{Sym}\int_{0}^{t}\mathrm{d}s_{3}\,
\int_{0}^{t}\mathrm{d}s_{1}\int_{0}^{s_{1}}\mathrm{d}s_{2}\,\mathds{K}_{i}(t-s_{3})\,
\tilde{\mathds{C}}_{i j}(s_{3},s_{2})\,\mathds{K}_{j}^{T}(s_{1}-s_{2})
\label{GT:def}
\end{eqnarray}
The {\color{black} superscript $T$ denotes} here and below the matrix transposition operation.

As a second step, we require that the correlation functions satisfy suitable 
integrability conditions. Specifically, we suppose that
\begin{hypothesis}{II}
\label{hyp:2}
\emph{there exists a positive-definite scalar function $F$ such that for any $t$, $t^{\prime}$}
\begin{eqnarray}
|\mathds{\tilde{C}}_{i j}^{m\,n}(t,t^{\prime})| < F(t-t^{\prime})\hspace{1.0cm}\forall\,m\,,n=1,\dots,d
\hspace{0.2cm}\&\hspace{0.2cm}\forall\,i\,,j=0,\dots,N
\nonumber
\end{eqnarray}
with $F(t)=F(-t)$ and
\begin{eqnarray}
\int_{0}^{\infty}\mathrm{d}t\,F(t)\,=f_{\star}<\infty
\nonumber
\end{eqnarray}
\end{hypothesis}
In words, we are hypothesizing that Lagrangian correlations decay sufficiently fast to take limits under the 
multiple integral sign. This is important because in appendix
we show that for any finite $t$
\begin{eqnarray}
\lefteqn{
\operatorname{Sym} \int_0^t      \mathrm{d}s\,     
\langle\,\delta\boldsymbol{v}(t)\otimes\delta\boldsymbol{v}(s)\rangle
=}
\nonumber\\
&&\sum_{i j=0}^{N}\operatorname{Sym}\int_{0}^{t}\mathrm{d}s_{3}\,
\int_{0}^{t}\mathrm{d}s_{1}\int_{0}^{s_{1}}\mathrm{d}s_{2}\,\mathds{K}_{i}(s_{3})\,
\tilde{\mathds{C}}_{i j}(t-s_{3},t-s_{2})\,\mathds{K}_{j}^{T}(s_{1})
\label{GT:adapted}
\end{eqnarray}
We thus set the scene to introduce our last hypothesis. We posit that
\begin{hypothesis}{III}
\label{hyp:3}
\emph{all the Lagrangian correlation functions (\ref{GT:nz}) have a well defined stationary limit }
\begin{eqnarray}          
\C_{i j}(t)=
\lim_{t^{\prime}\uparrow\infty }\tilde{\C}_{i j}(t+t^{\prime},t^{\prime})
\end{eqnarray}
\end{hypothesis}
An immediate consequence of the definition (\ref{GT:nz}) and of hypothesis~\textbf{III} is that for any finite $t$
\begin{eqnarray}
\C_{i j}(t)=\lim_{t^{\prime}\uparrow\infty}\tilde{\C}_{j i}^{T}(t^{\prime},t^{\prime}+t)=
\lim_{t^{\prime}\uparrow\infty}\tilde{\C}_{j i}^{T}(t^{\prime}-t,t^{\prime})=
\C_{j i}^{T}(-\,t)
\label{GT:property}
\end{eqnarray}
In appendix we combine hypotheses \textbf{I-II-III} to show that
 Taylor's identity holds true in the generalized form
 {\color{black}
\begin{eqnarray}
\label{GenTaylor} \D&= &
\sum_{i,j=0}^{N}\mathds{\hat{K}}_{i}(0) \,\frac{\mathds{\hat{C}}_{i j}(0)+\mathds{\hat{C}}_{j i}^{T}(0)}{2}\,\mathds{\hat{K}}_{j}^{T}(0)=\sum_{i,j=0}^{N} \mathds{\hat{K}}_{i}(0)  \int_0^\infty     \mathrm{d}s\,\frac{\mathds{{C}}_{i j}(s)+\mathds{{C}}_{j i}^{T}(s)}{2}  \,\mathds{\hat{K}}_{j}^{T}(0)
\end{eqnarray}
}
with
\begin{eqnarray}
\mathds{\hat{C}}_{i j}(z)=\int_0^\infty \mathrm{d}t \,e^{-z\,t}\,\C_{i j}(t)\,,\hspace{0.5cm}\operatorname{Re}z>0
\label{GT:CLaplace}
\end{eqnarray}
A few remarks on the nature of the hypotheses are in order. 

The validity of hypothesis~\textbf{I}
can be checked a priori from the explicit form of the equation of motion of inertial particle models.

Hypotheses~\textbf{II-III}  are instead not obviously 
granted. Their validity is an assumption on the properties of the solutions of (\ref{GT:eq}). From the
physics slant, we need hypothesis~\textbf{II} to control memory effects. For example, relaxation dynamics of
infinite dimensional systems with Boltzmann equilibrium may give rise to ageing phenomena \cite{Bir05}. 
Similar very slow decay of Lagrangian correlations must be ruled out in order to apply the dominated convergence 
theorem which we need to arrive at {\color{black} generalized} Taylor's formula.

Eulerian carrier velocity field and external forces are in general explicit functions of the 
time variable. Lagrangian correlation functions may become asymptotically stationary (hypothesis~\textbf{III}) 
in consequence of the ensemble average operation $\langle\dots\rangle$. For example, hypotheses~\textbf{II-III}  are satisfied 
if the Eulerian statistics of velocity field is a random Gaussian ensemble 
delta correlated in time, a widely applied stylized model of turbulent field \cite{FGV01}. 

Finally, the foregoing hypotheses are essentially the same as those underlying
the derivation of Green--Kubo formulas \cite{KuToHa91} in non-equilibrium statistical mechanics.
It is in this sense justified to regard {\color{black} generalized} Taylor's formula as the hydrodynamic version of these relations.

\section{Basset--Boussinesq--Oseen model}
\label{sec:BBO}

We now turn to apply the general results of section~\ref{sec:GT} to explicit models 
of dynamics. To  start with, let  us
consider the  simplest and oldest  model for inertial particles  in an
incompressible  flow, the  so-called  Basset--Boussinesq--Oseen equation \cite{Tch47}:
\begin{equation}\label{BBO}
\begin{split}        
\frac{\mathrm{d}\boldsymbol{v}}{\mathrm{d}t}(t)&=        
\frac{\boldsymbol{u}(\boldsymbol{\xi}(t),t)        -       \boldsymbol{v}(t)
}{\tau}+\beta     \frac{\mathrm{d}\boldsymbol{u}(\boldsymbol{\xi}(t),t)}{\mathrm{d}t}+
\boldsymbol{f} \\  & +  \sqrt{\frac{3\beta}{\pi
\tau}}\int_{0}^{t}\frac{\mathrm{d} s}{\sqrt{t    -s}}\frac{\mathrm{d}}{\mathrm{d}
s}\Big{(}\boldsymbol{u}(\boldsymbol{\xi}(s),s)                  -\boldsymbol{v}(s)\Big{)}
  \end{split}
\end{equation} 
In  the above equation, {\color{black} $\boldsymbol u$ is the undisturbed flow}, the pressure gradient  term is
estimated   as   $\boldsymbol\nabla  p \propto -   \mathrm{d}\boldsymbol   u/\mathrm{d}t$,
\cite{Tch47}       where      $\frac{\mathrm{d}}{\mathrm{d}t}=\partial_t+\boldsymbol
v\cdot\boldsymbol\nabla$;  the  term  $\boldsymbol  f$  is  a  generic
external force {\color{black} per unity of mass}
, {\color{black}$\tau\equiv r_p^2/(3\nu\beta)$ denotes the Stokes time,  $r_p$ being  the  radius  of inertial  particles } (supposed to  be
spherical) and $\nu$ the  fluid kinematic viscosity.  
Finally,  the  parameter
$\beta$  is  the  added-mass  factor,  $\beta\equiv  3\rho_f/(\rho_f+2
\rho_p))\in [0,3]$ built from the  constant fluid density $\rho_f$ and
the particle  density $\rho_p$.  {\color{black} Eq. (\ref{BBO}) assumes no-slip condition on the particle surfaces. It had been initially used to model a motion of a particle in a static and uniform flow $\boldsymbol u$. Afterwards, it has been proposed  for the dynamics of particles in a non-uniform and time-dependent flow too, under a number of approximation (\cite{Tch47}). Firstly, it describes very small particles, and any term  $\sim o(r_p/ L)$ is neglected, with L being the minimal variation length of the flow. Secondly, the Reynolds number with respect to the particle motion Re$_p=(\max |\boldsymbol u- \boldsymbol v|) r_p/\nu$ is supposed to be sufficiently close to 0. Finally, the Stokes number, that is  the ratio between Stokes time  and  the smallest advection time $\tau_F$ in the flow,  should be far smaller than 1, i.e. $\tau/\tau_F\ll1$. Under these approximations, Eq. (\ref{BBO}) represents the lowest order approximation with respect to these parameters towards the more modern Maxey-Riley model \cite{MaRi83}.} The  last integral in  (\ref{BBO}) is
the  Basset history  term  describing  the force  due  to the  lagging
boundary  layer development  with  changing relative  velocity of  the
particle  moving   through  the   fluid,  under  the   condition  that
$\boldsymbol      v(0)=\boldsymbol       u(\boldsymbol      \xi(0),0)$
\cite{Maxey1987}. {\color{black} If the latter is not satisfied, alternative forms for the history term are available in literature \cite{FaHa15}, and they in fact preserve the Laplace transform of Eq. (\ref{BBO}). However, this aspect does not affect our analysis, as only the Laplace transforms of the integral kernels $\mathds{\hat{K}}_{j}$ enter into  Eq. (\ref{GenTaylor}).

\subsection{The buoyancy-forced case}

\subsubsection{Constant force}

If $\boldsymbol  f$ represents the buoyancy  contribution described by
the     term     $(1-\beta)\boldsymbol    g$     \cite{MazMarMur2012}, the Fourier-Laplace
transform of Eq. (\ref{BBO}) yields
\begin{eqnarray}
\label{vs}  
\hat{\boldsymbol{v}}(z)
&=&
\frac{(1-\beta) \boldsymbol u(\boldsymbol \xi(0),0)}{a(z)}+\frac{(\beta-1)z+a(z)}{a(z)}\,\hat{\boldsymbol{u}}(z)
+\frac{(1-\beta)\,\boldsymbol{g}}{z\,a(z)}
\end{eqnarray} 
where
\begin{eqnarray}
\hat{\boldsymbol{v}}(z)\equiv\int_{0}^{\infty}\mathrm{d}t\,e^{-z\,t}\,\boldsymbol{v}(t)
\nonumber
\end{eqnarray}
and
\begin{eqnarray}
\hat{\boldsymbol{u}}(z)\equiv\int_{0}^{\infty}\mathrm{d}t\,
e^{-z\,t}\,\boldsymbol{u}(\boldsymbol{\xi}(t),t)
\nonumber
\end{eqnarray}
and finally
\begin{eqnarray}
\label{BBO:coe}
a(z)=z+\frac{1}{\tau}+ \sqrt{\frac{3\beta z}{\pi
\tau}}
\end{eqnarray}
If  we contrast Eq.  (\ref{vs}) with the Fourier-Laplace transform of Eq. (\ref{prima}) 
\begin{eqnarray}
\label{primas}  \hat{\boldsymbol{v}}(z)&=&
\boldsymbol{\hat\sigma}(\boldsymbol {\xi}(0),\boldsymbol v(0),z)+ \mathds{\hat{K}}_0(z)\,\hat{\boldsymbol{u}}(z)+\sum_{i=1}^{N} \mathds{\hat{K}}_{i}(z)\,
\boldsymbol{  \hat{f}}_{i}(z)\;, 
\end{eqnarray} 

we readily see that (\ref{vs}) corresponds to the case $N=1$ with 
\begin{eqnarray}
\boldsymbol{\hat{f}}_{1}(z)=\frac{(1-\beta)\,\boldsymbol{g}}{z}
\nonumber
\end{eqnarray}
and
\begin{eqnarray}
\mathds{\hat{K}}_{0}(z)&=&\hat{K}_0(z)\,\mathds{1}=\frac{(\beta-1)\,z+a(z)}{a(z)}\,\mathds{1}
\nonumber\\
\mathds{\hat{K}}_{1}(z)&=&\hat{K}_1(z)\,\mathds{1}=\frac{1}{a(z)}\,\mathds{1}
\nonumber\\
\boldsymbol{\hat\sigma}(\boldsymbol {\xi}(0),\boldsymbol v(0),z)&=&(1-\beta)\boldsymbol u(\boldsymbol\xi(0),0) \hat K_1(z)
\end{eqnarray}
satisfying $\hat K_0(0)=1 $ and $\hat K_1(0)=\tau$. 

We are now going to prove the transient does not play any role in asymptotic diffusion nor does it provide any dependence on the initial conditions. By virtue of the properties of Laplace transform, $\boldsymbol{\hat\sigma}(\boldsymbol {\xi}(0),\boldsymbol v(0),z)$ in physical space is equivalent to a convolution between $K_1(t)$ and the external forcing term:
\[
\boldsymbol f_2=(1-\beta)\,\boldsymbol u(\boldsymbol\xi(0),0)\,\delta(t)
\]
To verify the validity of hypothesis III for such term, we need to calculate the limits \begin{eqnarray}
\label{nulltransient}
\C_{22}(t)&=&\lim_{t^\prime\to\infty}\langle\delta\boldsymbol  f_2(t^\prime)\otimes\delta\boldsymbol f_2(t+t^\prime)\rangle=0\nonumber\\
\C_{21}(t)&=&\lim_{t^\prime\to\infty}\langle\delta\boldsymbol f_2(t^\prime)\otimes\delta\boldsymbol g\rangle=0\nonumber\\
\C_{20}(t)&=&\lim_{t^\prime\to\infty}\langle\delta\boldsymbol f_2(t^\prime)\otimes\delta\boldsymbol u(\boldsymbol\xi(t+t^\prime),t+t^\prime)\rangle=0
\end{eqnarray}
since $\delta (t^\prime)$ has no support for any $t^\prime>0$. In Eq. (\ref{nulltransient}) we indicated  $\delta \boldsymbol u= \boldsymbol u-\langle\boldsymbol u\rangle$ and clearly $\delta\boldsymbol g=\boldsymbol g-\langle\boldsymbol g\rangle=0$. We therefore conclude that the transient term originating from the initial condition does not contribute to asymptotic diffusion in the generalized Taylor formula for the Basset-Boussinesq-Oseen model and it will be ignored from now on. Neglecting the vanishing transient, the inverse Fourier--Laplace transform of Eq. (\ref{primas}) yields the explicit form of Eq. (\ref{prima}) 
\begin{eqnarray} 
\boldsymbol{v}(t)&=&
\int_0^t          \mathrm{d}s   \,       K_0(t-s)\,\boldsymbol{u}(s)
+\left(1-\beta\right)\,\boldsymbol{g}\,\int_0^ t\mathrm{d}s\, K_1(s)
\end{eqnarray} 
By virtue of Eq. (\ref{GenTaylor}) we obtain that the usual  Taylor 1921's  formula for tracer holds true for this case:
\begin{equation}
\D=\operatorname{Sym}\mathds{\hat{C}}_{0 0}(0)=\lim_{t\uparrow\infty }\operatorname{Sym} \int_0^t      \mathrm{d}s\,     
\langle\,\delta\boldsymbol{u}(\boldsymbol \xi(s),s) \otimes\delta\boldsymbol{u}(\boldsymbol \xi(t),t) \rangle \;\;\;,
\end{equation} 
and the trace of  the resulting  eddy-diffusivity tensor is the same that had been found in \cite{Tch47}}. 

\subsubsection{Brownian force}

We can repeat the same steps as above in the presence  of an external  
Brownian force {\color{black} per mass unity equal to}  $   \sqrt{2   D_0}/\tau\;    \boldsymbol\eta(t)$,
$\boldsymbol\eta(t)$ being a white-noise process coupled by a constant
molecular diffusivity $D_0$ \cite{Ree88}.  The equation for the particle
velocity can be couched (neglecting transient terms) into the form (\ref{prima}):
\begin{eqnarray} \boldsymbol{v}(t)&=&
\int_0^t \mathrm{d}s\,  K_0(t-s)\,\boldsymbol{u}(z)
+ \int_0^t \mathrm{d}s \,K_1(t-s)
\Big{(}
(1-\beta)\,\boldsymbol{ g}+\frac{\sqrt{2\,  D_0}}{\tau}\, 
\boldsymbol{\eta}(s)
\Big{)}
\end{eqnarray}
{\color{black} Here, we have the following expressions for $\boldsymbol\phi_i(\boldsymbol \xi(t),t)$:
\begin{equation}
\boldsymbol\phi_i(\boldsymbol \xi(t),t)=\left\{
\begin{aligned}
&\boldsymbol u(\boldsymbol \xi(t),t) -\Big{\langle} \boldsymbol u(\boldsymbol \xi(t),t)\Big{\rangle}\quad &\text{if }i=0&\\
&\frac{\sqrt{2\,  D_0}}{\tau}\, 
\boldsymbol{\eta}(s)\quad &\text{if }i=1&\\
\end{aligned}\right.
\end{equation}
As a result of this:
\begin{equation}
\left\{
\begin{aligned}
\C_{00}(t)&=\lim_{t^{\prime}\uparrow\infty } \tilde{\C}_{00}(t+t^{\prime},t^{\prime})=\lim_{t^{\prime}\uparrow\infty }        
\langle\,\delta\boldsymbol{u}(\boldsymbol \xi(t+t^{\prime}),t+t^{\prime})\otimes\delta\boldsymbol{u}(\boldsymbol\xi(t^{\prime}),t^{\prime})\rangle&\\
\C_{11}(t)&=\lim_{t^{\prime}\uparrow\infty } \tilde{\C}_{11}(t+t^{\prime},t^{\prime})=\frac{2 D_0}{\tau^2}\,\delta(t)&\\
\C_{10}(t)&=\C_{02}(t)=0&\\
\end{aligned}\right.
\end{equation}
the last equality being a consequence of causal independence between white noise $\boldsymbol{\eta}(t)$ and fluid velocity $\boldsymbol{u}(t^\prime)$ at time $t^\prime\leq t$. 
Upon recalling the identity $\int_0^\infty \mathrm{d}t \delta(t)=1/2$ and Eq. (\ref{GenTaylor}), } we arrive at the 
expression of the eddy diffusivity:
\begin{equation}
\label{TaylorBBO} \D=D_0\,\mathds{1}+ 
\operatorname{Sym}\mathds{\hat{C}}_{0 0}(0)
\end{equation} 
which is formally the same Lagrangian expression as that
of  tracers. This does  not  mean at  all  that the  eddy
diffusivity  of  tracers  and  of   inertial  particles  must  be  the
same.  Eq. (\ref{TaylorBBO})  is indeed  evaluated along  trajectories
which differ in the two cases.
                                                                                                                                                   
\subsection{Inclusion of the Lorentz force}

{\color{black}A generalized form of Taylor's formula is possible} if inertial particles are subject to 
a  Lorentz   force  $-q\boldsymbol
B\times\boldsymbol  v$ in  a constant  magnetic field  $\boldsymbol B$ and
inter-particle interactions are neglected \cite{Guazzelli}. This can be regarded as a stylized model 
of charged particles in a plasma \cite{Kur62,LeKa99,CzGa01}.  Furthermore,  it is possible to show
that when in a solid the electron-electron collision mean-free path is
far smaller than the system width,  electrons can be modeled as a
fluid  where  mutual collisions  are  taken  into account  by  viscous
dissipation \cite{Ale16}.

The Laplace transform of the equation of motion without transient yields: 
\begin{eqnarray}
\label{Lorentz}     &&
\mathds{\hat{A}}(z)\,\boldsymbol{\hat{v}}(z)=
\big{(}(\beta-1)\,z+a(z)\big{)}\,\boldsymbol{\hat{u}}(z)+\frac{\sqrt{2\,D_0}}{\tau}\boldsymbol{\hat{\eta}}(z)
\end{eqnarray} 
where we defined the strictly positive definite tensor $\mathds{\hat{A}}(z)$ with components
\begin{eqnarray}
\label{}
\mathds{\hat{A}}^{\mu\nu}(z)=
a(z)\,\delta^{\mu\nu}
&+&
\gamma\, B^\nu \epsilon^{\mu \sigma \nu}
\nonumber
\end{eqnarray}
where $a$  is defined by (\ref{BBO:coe}) and
\begin{eqnarray}
\label{BBO:coe3}
\gamma=\frac{q}{\frac{4}{3}    \,  \pi\, r_p^3 \,\rho_p }
\end{eqnarray}
Upon inverting $\mathds{\hat{A}}(z)$ we obtain an equation of the form (\ref{primas}), whence
it is straightforward to derive {\color{black} generalized} Taylor's formula
\begin{equation} 
\label{TaylorLorentz}
\D=\frac{D_0}{\tau^2} \mathds{\hat{A}}^{-1}(0)\,
(\mathds{\hat{A}}^{-1})^{T}(0)+  \frac{1}{\tau^2}  \mathds{\hat{A}}^{-1}(0) 
\operatorname{Sym}\Big{(}\hat{\C}_{0 0} (0)\Big{)} \, (\mathds{\hat{A}}^{-1})^{T}(0)
\end{equation}
with
\begin{eqnarray}
(\mathds{\hat{A}}^{-1})^{\mu\nu}(z)=\frac{1}{a^{2}(z)+\gamma^{2}\,\|\boldsymbol{B}\|^{2}}
\left[a(z)\,\delta^{\mu\nu}-\gamma\,B^{i}\,\epsilon^{\mu \sigma \nu}+\frac{\gamma^{2}}{a(z)}B^{k}\,B^{j}\right]
\nonumber
\end{eqnarray}
{\color{black} Notice that due to the Laplace transform on Eq. (\ref{Lorentz}), the transformed Green function $(\mathds{\hat{A}}^{-1})^{\mu\nu}(z)$ is dimensionally a time, and consistently Eq. (\ref{TaylorLorentz}) has the same dimensions of $D_0$. }

\subsubsection{Limit of vanishing carrier velocity field}

A  simple  application  is  when $\boldsymbol  u=0$  in $d=3$ and the
magnetic field $\boldsymbol B$ is oriented along the third coordinate axis ($\boldsymbol{B}=B\,\boldsymbol{e}_{3}$ for $\boldsymbol{e}_{3}$ is the unit vector spanning the axis). 
We get:
\begin{equation}
\label{camposolo}
\D=\text{diag}\left(\frac{D_0}{1+
\gamma^{2}\,B^2\,\tau^2},\frac{D_0}{1+
\gamma^{2}\, B^2 \,\tau^2},D_0\right)
\end{equation} 
where we  can observe a reduction of  the transport due
to    the   action    of    the   magnetic    field.   
Eq. (\ref{camposolo})   generalizes the result of \cite{LeKa99}, {\color{black} by
showing that the added mass effect and the Basset history term do not play any role in the asymptotic transport when the flow is at rest and a Lorenz force is present. This result is also in agreement with \cite{Russel,Reichl}, where it is shown that in still fluids Stokes drag term and Basset force create noise with memory which however has not effect on the eddy diffusivity. 
On the other hand,  there is much investigation in literature about strong  differences Basset history term can make in particle motion when the flow is not at rest. One of the most representative cases is \cite{Dru}. Therein, it is shown that in a cell flow heavy particles with small $\tau$ remain trapped into cells (i.e. no diffusion), whereas Basset history force term lets them escape along the cell separatrixes, resulting in oscillating ballistic trajectories. The latter effect gives rise to a infinite eddy diffusivity, i. e. superdiffusion \cite{BoiMazzMar}.} 

\subsubsection{Limit of vanishing Stokes number}

Another noteworthy case  is when the Stokes time $\tau$  is much smaller
than the typical flow time scale $\tau_{F}$ (i.e. $\mathrm{St}\ll1,$ with $\mathrm{St}$ the
Stokes number  $\tau/\tau_{F}$) but $
\gamma\,\boldsymbol{B} \, \tau$  is  independent   of  $\tau$.  By  introducing  the
dimensionless  magnetic field
  $ \boldsymbol{B}^*=
\gamma\,\tau \,\boldsymbol{B}
$, 
Eq. (\ref{Lorentz}) becomes:
\begin{eqnarray}
\label{LorentzTracs}      
\mathds{A}\,\boldsymbol{\hat{v}}(z)=\boldsymbol{\hat{u}}(z)
+\sqrt{2\,D_0}\,\boldsymbol{\hat{\eta}}(z)
\end{eqnarray} 
with
\begin{eqnarray} 
\label{A}
&& \mathds{A}^{\mu\nu}=\delta^{\mu\nu}+B^{*\,
i}\,\epsilon^{\mu \sigma \nu}\;,
\end{eqnarray} 
Upon inverting the Laplace transform, the equation for the particle velocity is 
\begin{equation}
\label{LorentzTrac}      \frac{\mathrm{d}\boldsymbol{\xi}}{\mathrm{d}t}(t)=      
\mathds{A}^{-1}\,\boldsymbol{u}(\boldsymbol{\xi}(t),t)+\sqrt{2\, D_0} \, \mathds{A}^{-1}\,\boldsymbol{\eta}(t)
\end{equation}  
The  system is  equivalent  to  a  tracer advected by a compressible drift field  
  $     \tilde    {\boldsymbol{u}}=    \mathds{A}^{-1}\,\boldsymbol{u}   $  and subject to an anisotropic diffusion coefficient  $\boldsymbol \tilde{\sigma}=\sqrt{2 \, D_0}
\,\mathds {A}^{-1}$. 
The eddy diffusivity is in this case
\begin{equation} \D=D_0\; \mathds{A}^{-1} ({\mathds A}^{-1})^{T}+
\mathds{A}^{-1}   \operatorname{Sym}\Big{(}\hat{\C}_{0 0}(0)\Big{)}   ({\mathds
A}^{-1})^{T}
\end{equation} 
The limit of $\boldsymbol{B}^*\to  0$ then recovers Taylor's formula for tracer particles. {\color{black} Notice that now $\mathds{A}^{\mu\nu}$ is dimensionless by definition in Eq. (\ref{A}).}

\subsection{Inclusion of the Coriolis force}

The inclusion of Coriolis force in the Basset--Boussinesq--Oseen model in the geostrophic approximation limit and neglecting the history-force term 
 yields \cite{MoHeGaRoLo17}:
\begin{equation}\label{BBOCoriolis}
\begin{split}        
\frac{\mathrm{d}\boldsymbol{v}}{\mathrm{d}t}(t)&=        \frac{
\boldsymbol{u}(\boldsymbol{\xi}(t),t)-       \boldsymbol{v}(t)
}{\tau}+\beta     \frac{\mathrm{d}\boldsymbol    u(\boldsymbol{\xi}(t),t)}{\mathrm{d}t}+ (1-\beta)      \boldsymbol      g\nonumber\\
&- 2\boldsymbol\Omega\times\Big{(}\boldsymbol
v(t)-\beta\boldsymbol     u(\boldsymbol{\xi}(t),t)\Big{)}+\frac{\sqrt{2\, D_0}}{\tau}\boldsymbol{\eta}(t)   
\end{split}
\end{equation}  
According to the geostrophic approximation, the centrifugal force is a small constant term  
which can be absorbed in a re-definition of $\boldsymbol g$. The Fourier--Laplace transform of (\ref{BBOCoriolis}) yields
\begin{eqnarray}
\label{Coriolis} 
\mathds{\hat{A}}^{\mu\nu}(z)\,\hat{v}^\nu(z)
=\mathds{\hat{B}}^{\mu\nu}(z)\,\hat{u}^{\nu}(z)+\frac{1-\beta}{\tau}g^\mu
+\frac{\sqrt{2\,D_0}}{\tau}\,\hat{\eta}^{\mu}(z)
\end{eqnarray}
where we define
\begin{eqnarray}  &&\mathds{\hat{A}}^{\mu\nu}(z)=
\left(z+\frac{1}{\tau}\right)\,\delta^{\mu\nu}
+2\,\Omega^{\sigma}\,      \epsilon^{\mu \sigma \nu}
\nonumber\\ && 
\mathds{\hat{B}}^{\mu\nu}(z)=
\left(\beta\,z+\frac{1}{\tau}\right)\,\delta^{\mu\nu}
+2\,\beta\,\Omega^{\sigma}\, \epsilon^{\mu \sigma \nu}
\nonumber
\end{eqnarray} 
In (\ref{Coriolis}) and in other occasions below, we use the Einstein convention for repeated 
indexes labeling tensor spatial components. {\color{black} Generalized} Taylor's formula is in this case:
\begin{equation}   \D=\frac{D_0}{\tau^2}  
\mathds{\hat{A}}^{-1}(0)\,({\mathds{\hat{A}}}^{-1})^{T}(0)+ 
\mathds{\hat{A}}^{-1}(0)\,\mathds{\hat{B}}(0)\, 
\operatorname{Sym}\Big{(}\hat{\C}_{0 0}(0)\Big{)}   \,\mathds{\hat{B}}^T(0)\, ({\mathds{\hat{A}}}^{-1})^{T}(0)
\end{equation}

\subsubsection{Limit of vanishing carrier velocity field}

If we consider a situation of  zero flow, then the diffusion is caused
only by  the molecular  white noise.  We, thus, recover (\ref{camposolo})
with $\gamma=2$ and $B=\|\Omega\|$.
\subsubsection{Limit of vanishing Stokes number at fixed Rossby}

It is again worth to consider the limit of
small Stokes  time $\tau$ with respect  to the flow time  scale, whilst
holding fixed {\color{black} the Rossby} number $\mathrm{Ro}=1/(\tau\,\Omega)$. 

\begin{figure}
\includegraphics[width=13cm, height=5cm]{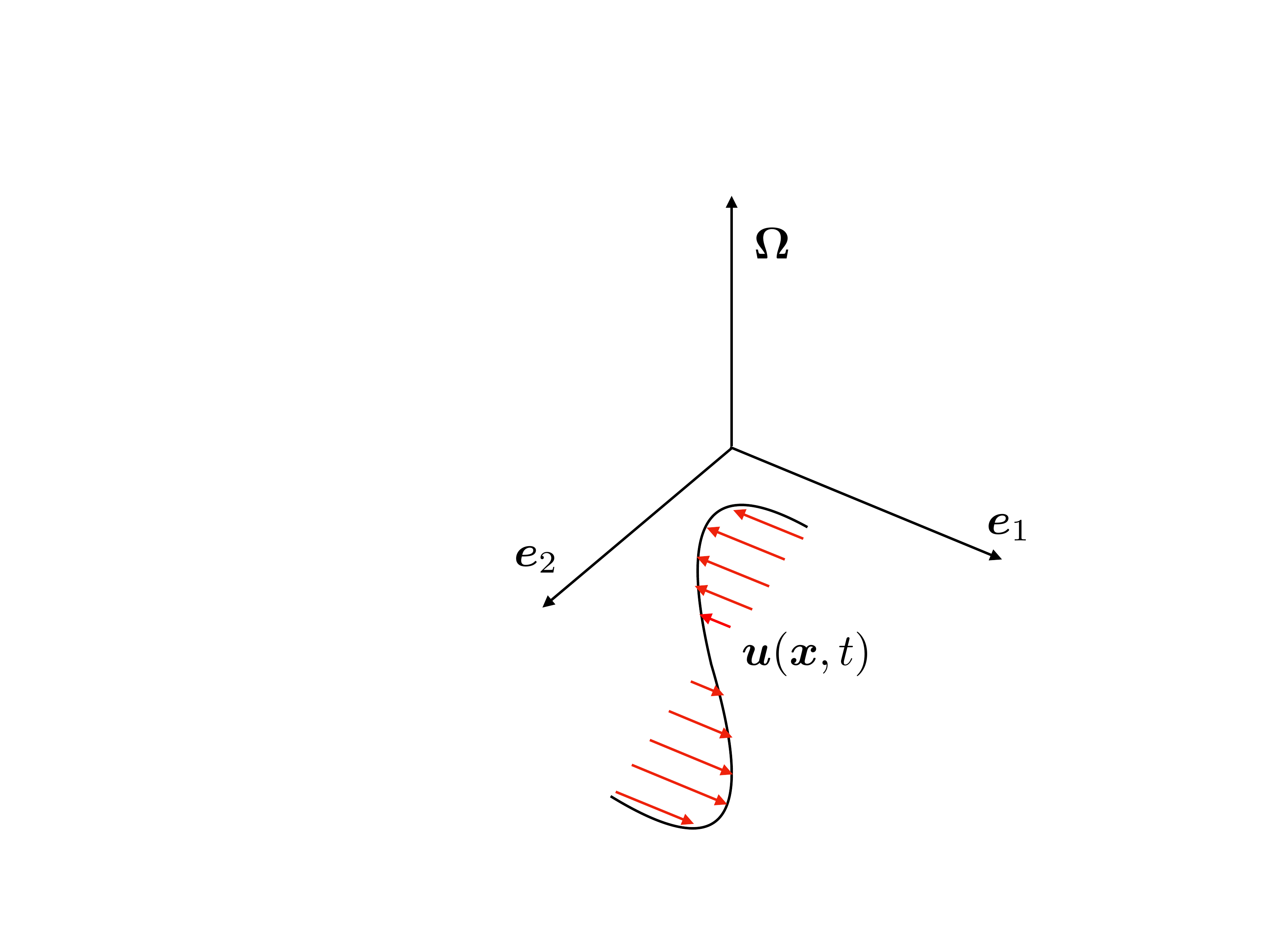}
\caption{Sketch of a shear flow along the direction $\boldsymbol e_1$ in a reference frame with angular velocity $\boldsymbol\Omega$ along $\boldsymbol e_3 $.}
\label{Fig1}
\end{figure}

If we define the constant matrices
\begin{eqnarray}  &&  
\mathds{A}^{\mu\nu}=\delta^{\mu\nu}+2\,\tau\,\Omega^{\sigma}\, \epsilon^{\mu \sigma \nu}
\nonumber\\  
&& \mathds{B}^{\mu\nu}=\delta^{\mu\nu}+2\,\beta\,\tau\,  \Omega^{\sigma}\, \epsilon^{\mu \sigma \nu}
\nonumber
\end{eqnarray}  
we can write the equation for the particle velocity as
\begin{equation}
\label{CoriolisTrac}    
\frac{\mathrm{d}\boldsymbol{\xi}}{\mathrm{d}t}(t)=     
\mathds{A}^{-1}\,\mathds{B}\,\boldsymbol{u}(\boldsymbol \xi(t),t)
+\sqrt{2\,D_0}\, \mathds{A}^{-1}\, \boldsymbol{\eta}(t)
\end{equation} 
The same  considerations apply here as
for Eq. (\ref{LorentzTrac}).  The eddy diffusivity becomes:
\begin{equation}
\label{TaylorCoriolis}   \D=
D_0\,    \mathds{A}^{-1}\,({\mathds
A}^{-1})^{T}+ \mathds{A}^{-1}\,\mathds{B}\, 
\operatorname{Sym}\Big{(}\hat{\C}_{0 0}(0)\Big{)} \mathds{B}^T\, ({\mathds A}^{-1})^{T}
\end{equation}
In order to illustrate the relative importance of the distinct contributions to this 
formula, it is expedient to consider a simple {\color{black} three dimensional model consisting of a shear flow on a rotating plane (see Fig. \ref{Fig1}). The angular velocity $\boldsymbol\Omega$ is oriented along the third axis $\boldsymbol e_3$ and the randomly fluctuating shear flow is:}
\begin{equation}
\label{Couette} \boldsymbol u(\boldsymbol{x}, t)= u(x_{2}, x_{3},t) \boldsymbol
e_{1}\; ,
\end{equation} 
with $\boldsymbol{e}_{1}$ being  the unit vector along the first coordinate axis. 

Under these hypotheses the tensor $\C_{0 0}(t)$ in (\ref{TaylorCoriolis}) has only one non vanishing 
component: $\C_{0 0}^{1 1}(t)=C(t)$. Thus, upon introducing the vector 
\[ \boldsymbol{M}=\mathds A^{-1}\,\mathds B\,\boldsymbol{e}_{x}= \frac{1}{1+4/\Ro^2}\left(\begin{aligned}  &  1+4\beta/\Ro^2  \\  &
2(\beta-1)/\Ro \\ & 0 \\
\end{aligned}\right)
\] 
{\color{black} Generalized} Taylor's formula takes the form
\begin{equation}
\label{EddyCoriolis}   \D=\frac{D_0}{1+4/\Ro^2}\mathds    1+   \boldsymbol
M\otimes\boldsymbol M\int_0^\infty\mathrm{d} t\, C(t)\;
\end{equation} 
or, componentwise:
\begin{eqnarray}
\label{CoriolisRo}            
\lefteqn{
\mathds{D}=\frac{1}{1+4/\mathrm{\Ro}^2}\times
}
\nonumber\\
&&
\begin{bmatrix}
D_0+\frac{(1+4\beta/\Ro^2)^2}{1+4/\Ro^2}\int_0^\infty  \mathrm{d}t\, C(t)  &
\frac{2(1+4\beta/\Ro^2)(\beta-1)/\Ro}{1+4/\Ro^2}\int_0^\infty       \mathrm{d}t\,
C(t)                   &                   0                   \\
\frac{2(1+4\beta/\Ro^2)(\beta-1)/\Ro}{1+4/\Ro^2}\int_0^\infty       \mathrm{d}t\,
C(t) &  D_0+ \frac{4(\beta-1)^2/\Ro^2}{1+4/\Ro^2}\int_0^\infty \mathrm{d}t\,
C(t) & 0 \\ 0 & 0 & D_0(1+4/\Ro^2)\\
\end{bmatrix}
\nonumber
\end{eqnarray} 
It is instructive to analyze the  behavior of the
trace of the eddy diffusivity as a function of the Rossby number $\Ro$:
\[                       \operatorname{Tr}\D=D_0\left(1+\frac{2}{(1+4/\Ro^2)}\right)+
\frac{(1+4\,\beta/\Ro^2)^2+4(\beta-1)^2/\Ro^2}{(1+4/\Ro^2)^2}\int_0^\infty
\mathrm{d}t\, C(t)
\] 
{\color{black} As a function of $\mathrm{Ro}$, $\operatorname{Tr}\mathds{D}$ 
turns out to be  monotonic, at fixed $\beta$ and $D_0$. Indeed, its first derivative is :
\[
-\frac{(\beta^2-1)\int_0^\infty
\mathrm{d}t\, C(t)-2 D_0 }{(\Ro^2+4)^2}8\ \Ro
\]
which has a constant sign, given that $\Ro \geq0$.} In the limit of vanishing Rossby number  (i.e.  ideally an infinite  value of $\Omega$), we get 
\begin{eqnarray}
\lim_{\mathrm{Ro}\downarrow 0}\operatorname{Tr}\mathds{D}=D_0+\beta^2 \int_0^\infty \mathrm{d} t\, C(t)
\nonumber
\end{eqnarray}
The opposite limit of large Rossby (i.e.  the absence of rotation),  recovers 
the expression of the tracer particle model
\begin{eqnarray}
\lim_{\mathrm{Ro}\uparrow \infty}\operatorname{Tr}\mathds{D}=3\,  D_0+\int_0^\infty \mathrm{d} t\, C(t)
\nonumber
\end{eqnarray}
For $\beta<1$, $\operatorname{Tr}\mathds{D}$ always grows with respect to Ro.  
For light particles ($\beta>1$), instead, the $\operatorname{Tr}\mathds{D}$ may be monotonically 
decreasing or  increasing depending upon whether the diffusion
contribution   from  the   flow   $\int_0^\infty  \mathrm{d} t\,  C(t)$   is
respectively higher or lower than the threshold value:
\[ \frac{2 D_0 }{\beta^2-1}.
\]  
For incompressible carrier fields  $\boldsymbol u$, 
$\int_0^{\infty}\mathrm{d}t\, C(t)$ is always positive. Hence, it is clear that only
for $\beta>1$ a decreasing behavior is possible.

\section{Maxey-Riley model}
\label{sec:MR}

We now turn to the derivation of {\color{black} generalized} Taylor's formula for the now {\color{black}``canonical''  Maxey--Riley 
model \cite{MaRi83} inclusive of the time derivatives along fluid trajectories and the Fax\'en friction \cite{Auton,AuHuPr88}:}
\begin{equation}\label{MR}
\begin{split}        
\frac{\mathrm{d}\boldsymbol         v}{\mathrm{d}t}(t)&=        \frac{
\boldsymbol{u}(\boldsymbol{\xi}(t),t) - \boldsymbol{v}(t) +\frac{1}{6}
r_p^2   \nabla^2   \boldsymbol   u(\boldsymbol{\xi}(t),t)}{\tau}+\beta
\frac{\mathrm{D}\boldsymbol   u(\boldsymbol{\xi}(t),t)}{\mathrm{D}t}+   \frac{\beta}{30}
r_p^2 \frac{\mathrm{d} }{\mathrm{d}t}\nabla^2  \boldsymbol u(\boldsymbol{\xi}(t),t) \\ &
+ \sqrt{\frac{3\beta}{\pi
\tau}}\int_{0}^{t}\frac{\mathrm{d}s}{\sqrt{t-s}}\,
\frac{\mathrm{d}}{\mathrm{d}s}
\Big{(}\boldsymbol{u}(\boldsymbol{\xi}(s),s) -\boldsymbol{v}(s)+\frac{1}{6}
r_p^2   \nabla^2   \boldsymbol{u}(\boldsymbol{\xi}(s),s)\Big{)}
+\frac{\sqrt{2D_0}}{\tau} \boldsymbol{\eta}(t)
\end{split}
\end{equation} 
{\color{black}where $\frac{D}{Dt}=\partial_t+\boldsymbol
 u \cdot\boldsymbol\nabla$.  With respect to the Basset-Boussinesq-Oseen equation, the latter term represents a higher order correction in the Stokes number, which still needs to be small \cite{MaRi83}. Higher order corrections in particles size are included, thanks to the Fax\'en drag force \cite{Gatignol}. The reason is to take into account terms of order O($r^2_p/L^2)$ whenever they could produce small but relevant deviations in comparison to the lower order approximation the Stokes drag provides.  These higher-order corrections with respect to Stokes number and particle radius are often included in applications \cite{ToBo09, Guazzelli}. } For simplicity sake, we do not discuss here external forces.   
Upon removing the initial transient, taking  Fourier--Laplace transform and recalling (\ref{BBO:coe}), we get into

\begin{eqnarray}
\label{MRs} 
\lefteqn{              
\boldsymbol{\hat{v}}(z)=
\frac{1}{a(z)}\left[
\beta\widehat{\frac{\mathrm{D}\boldsymbol u}{\mathrm{D}t}}(z)
+\left(a(z)-z-\frac{1}{\tau}\right)\left(
\hat{\boldsymbol  u}(z)+\frac{1}{6} r_p^2 \widehat{\nabla^2\,\boldsymbol{u}}(z)\right)\right]
}
\nonumber\\
&+&\frac{1}{\tau\,a(z)}\left[\hat{\boldsymbol  u}(z)+\frac{1}{6} r_p^2
\widehat{\nabla^2\,\boldsymbol{u}}(z)+\sqrt{2   D_0}\,\boldsymbol{\hat{\eta}}(z)
+ z\,\frac{\beta}{30} \, r_p^2\, \widehat{\nabla^2
\boldsymbol {u}}(z)\right]
\end{eqnarray} 
Again the model can be couched into the form (\ref{prima}) with
all tensors $\mathds{K}_{i}$'s having the form of the identity matrix times scalar functions $K_{i}$ $i=0,\dots,3$. 
By comparing to Eqs. (\ref{primas}), we see that 
\begin{equation}
\label{MRkernels}
\begin{aligned}        
&\hat{K}_0(z)=1-\frac{z}{a(z)}
&\;                \\           
&\hat{K}_1(z)=\frac{1}{\,a(z)}              
&\hspace{0.2cm}\&\hspace{0.5cm}&              
\hat{\boldsymbol f}_1(z)=\beta\,\widehat{\frac{\mathrm{D}\boldsymbol{u}}{\mathrm{D}t}}(z)
+\frac{\sqrt{2\,D_0}}{\tau}\,\boldsymbol{\hat{\eta}}(z)
\\                 
&\hat{K}_2(z)= \tau\, \hat{K}_0(z)
&\hspace{0.2cm}\&\hspace{0.5cm}&
\boldsymbol{\hat{ f}}_2(z)=\frac{1}{6\tau}  r_p^2
\widehat{\nabla^2          \boldsymbol{u}}(z)
\\          
&\hat{K}_3(z)=\frac{\tau\,z}{a(z)}  &\hspace{0.2cm}\&\hspace{0.5cm}&
\boldsymbol{\hat{ f}}_3(z)=  \frac{\beta}{30\tau}\, r_p^2 \widehat{\nabla^2 \boldsymbol{u}}(z)
\end{aligned}
\end{equation} 
The scalar kernels satisfy
\begin{eqnarray}
\hat  K_0(0)&=&\int_0^\infty\mathrm{d}t\,
K_0(t)=1\nonumber\\
\hat  K_i(0)&=&\int_0^\infty\mathrm{d}t\,
K_i(t)=\tau\,\hspace{1.0cm}i=1,2
\nonumber
\end{eqnarray} 
whilst $\hat K_3(0)=0$.  This latter fact implies  that
$\boldsymbol f_3$  does not  give any  contribution to {\color{black} generalized} Taylor's formula. Upon applying the general result (\ref{GenTaylor}), 
we get into
{\color{black}
 \begin{eqnarray}
 \label{MRTaylor}      \D=D_0\mathds     1&+&\operatorname{Sym}\int_{t_0}^\infty      \mathrm{d}t\,
\left\langle\left[\delta \boldsymbol{u}(\boldsymbol{\xi}(t),t)
+\beta\tau\,\delta\frac{\mathrm{D}\boldsymbol{u}}{\mathrm{D}t}(\boldsymbol{\xi}(t),t) 
+\frac{1}{6}  r_p^2 \, \delta\nabla^2 \boldsymbol{u}(\boldsymbol{\xi}(t),t)\right]\right.
\nonumber\\
&\otimes&\left.\left[\delta \boldsymbol{u}(\boldsymbol{\xi}(t_0),t_0)
+\beta\tau\,\delta\frac{\mathrm{D}\boldsymbol{u}}{\mathrm{D}t}(\boldsymbol{\xi}(t_0),t_0)    
+\frac{1}{6}    r_p^2  \,\,\delta  \nabla^2 \boldsymbol{u}(\boldsymbol{\xi}(t_0),t_0)
\right]\right\rangle
 \end{eqnarray} 
where we  considered
the instant $t_0$  as the time at which the  correlation functions can
be  considered  stationary, and:
\begin{eqnarray}
\delta\boldsymbol u(\boldsymbol \xi(t),t) &=&\boldsymbol u(\boldsymbol \xi(t),t) -\Big{\langle} \boldsymbol u(\boldsymbol \xi(t),t)\Big{\rangle}\nonumber\\
\delta\frac{\mathrm{D}\boldsymbol u }{\mathrm{D}t}(\boldsymbol \xi(t),t)&=&\frac{\mathrm{D}\boldsymbol u}{\mathrm{D}t}(\boldsymbol \xi(t),t)  -\Big{\langle} \frac{\mathrm{D}\boldsymbol u}{\mathrm{D}t}(\boldsymbol \xi(t),t) \Big{\rangle}\\
\delta  \nabla^2\boldsymbol u(\boldsymbol \xi(t),t)& =& \nabla^2\boldsymbol u(\boldsymbol \xi(t),t) -\Big{\langle}  \nabla^2\boldsymbol u(\boldsymbol \xi(t),t)\Big{\rangle}\nonumber
\end{eqnarray}

}

By comparing  Eqs.  (\ref{MRTaylor})  and
(\ref{TaylorBBO}),  we clearly  see  the  Maxey--Riley and  Basset--Boussinesq--Oseen  models tend  to
coincide when  $r_p/L$ and $\beta\tau/\tau_{F}$  are $\ll  1$, $\tau_{F}$ and  $L$ being
characteristic time and length scale  of the flow, respectively.

Eq. (\ref{MRTaylor}) generalizes results previously given in  literature 
(see e.g.\cite{BoMaMG17}), {\color{black} where explicit expressions for the eddy diffusivity had been derived in the case of heavy particles. Indeed, that corresponds to $\beta=0$, and in such a limit only the Stokes drag in Eq. (\ref{MR}) survives.}

{\color{black} \section{Models including lift forces}
Further higher-order corrections due to particle size and higher Stokes numbers include lift forces. Some models were obtained in literature even in the case of small particle Reynolds numbers Re$_p$. The earliest model was provided by Saffman in 1965 \cite{Saffman,SaffmanII} for small solid particles in shear flows. This model is often used in its generalization to 3-dimensional flows \cite{Lift}. A lot of different, empirical models have been proposed since then, taking into account different sizes and shapes of particles, wall effects,  momentum transfer between the carrier fluid and the inner fluid inside the particle -- which is meaningful if that particle is a bubble -- or finite Reynolds numbers \cite{Legendre, LiftWall, Tomi, MeiLift}. Typically, these models have the following shape for the lift force on a spherical particle:
\begin{equation}
\label{LiftForce}
\boldsymbol F_L={C_L}\, \rho_f\, \frac{4}{3}\pi r_p^3 [\boldsymbol v(t)-\boldsymbol u(\boldsymbol\xi(t),t)]\times\boldsymbol\omega(\boldsymbol\xi(t),t)
\end{equation}
where $\boldsymbol\omega=\nabla\times\boldsymbol u$ is the vorticity and $C_L$ is the lift coefficient, which in general can be determined solely by fitting experimental data and it depends on several parameters of the carrier flow itself. 

It is not in the scope of this article to provide a general view over lift force models, which is a vast phenomenology as said above. We rather want to provide an example about how to obtain an expression of the eddy diffusivity via the generalized Taylor's formula. That would be useful to see how the autocorrelation and the mutual correlations of the several forces would act on the asymptotic diffusion. To do so, we stick to the Saffman model, for which \cite{Lift}:
\begin{equation}
C_L=\frac{6.46}{\frac{4}{3}\,  \pi r_p}\sqrt{\frac{\nu}{||\boldsymbol\omega(\boldsymbol\xi(t),t)||}}
\end{equation}
The equation of motion turns out to be:
\begin{equation}\label{MRSaffman}
\begin{split}        
\frac{\mathrm{d}\boldsymbol         v}{\mathrm{d}t}(t)&=        \frac{
\boldsymbol{u}(\boldsymbol{\xi}(t),t) - \boldsymbol{v}(t) +\frac{1}{6}
r_p^2   \nabla^2   \boldsymbol   u(\boldsymbol{\xi}(t),t)}{\tau}+\beta
\frac{\mathrm{D}\boldsymbol   u(\boldsymbol{\xi}(t),t)}{\mathrm{D}t}+   \frac{\beta}{30}
r_p^2 \frac{\mathrm{d} }{\mathrm{d}t}\nabla^2  \boldsymbol u(\boldsymbol{\xi}(t),t) \\ &
+ \sqrt{\frac{3\beta}{\pi
\tau}}\int_{0}^{t}\frac{\mathrm{d}s}{\sqrt{t-s}}\,
\frac{\mathrm{d}}{\mathrm{d}s}
\Big{(}\boldsymbol{u}(\boldsymbol{\xi}(s),s) -\boldsymbol{v}(s)+\frac{1}{6}
r_p^2   \nabla^2   \boldsymbol{u}(\boldsymbol{\xi}(s),s)\Big{)}\\
&+\frac{6.46\,\beta}{ 2\,\pi r_p}\sqrt{\frac{\nu}{||\boldsymbol\omega(\boldsymbol\xi(t),t)||}} [\boldsymbol v(t)-\boldsymbol u(\boldsymbol\xi(t),t)]\times\boldsymbol\omega(\boldsymbol\xi(t),t)+\frac{\sqrt{2D_0}}{\tau} \boldsymbol{\eta}(t)\\
\end{split}
\end{equation} 
Seeing that here the advection time scales are provided by the very vorticity, i.e. $\tau_F=\max(1/||\boldsymbol\omega||)$, and regcalling that $\tau= 3 r^2_p/(\nu\beta)$, the ratio between Saffman force per unity of mass and Stokes drag is:
\begin{equation}
\frac{\frac{2}{3}\beta \frac{6.46}{ 4/3\,\pi r_p}\sqrt{\frac{\nu}{||\boldsymbol\omega||}} ||[\boldsymbol v-\boldsymbol u]\times\boldsymbol\omega||}{|| \boldsymbol{u}       -       \boldsymbol{v}||/\tau}\leq \frac{2}{3}\beta  \frac{6.46}{4/3\pi}\sqrt{\frac{\nu}{||\boldsymbol\omega||r_p^2}}\tau||\boldsymbol\omega||\sim O(\sqrt\text{St})
\end{equation}
As a result of this, Saffman's lift force is always negligible at sufficiently low Stokes times, or whenever $\rho_p\ll\rho_f$, that is $\beta\ll1.$ For the Saffman model to hold true, along with Re$_p\sim0$ one needs: 
\[
\max\frac{||\boldsymbol\Omega_p||r_p^2}{||\boldsymbol v-\boldsymbol u||}\ll1\qquad\&\qquad\max\frac{\sqrt{||\boldsymbol \omega||/\nu}}{||\boldsymbol v-\boldsymbol u||/\nu}\gg1
\]
having indicated the particle angular velocity by $\boldsymbol \Omega_p$. 

We observe that we have one more forcing term in addition to those of the Maxey-Riley model in Eq. (\ref{MRkernels}):
\begin{equation}
\begin{aligned}        
&\hat{K}_4(z)=\hat{K}_1(z)  &\hspace{0.2cm}\&\hspace{0.5cm}&
\boldsymbol{\hat{ f}}_4(z)=  {\hat{\boldsymbol f}_L(z)}
\end{aligned}
\end{equation} 
where $\hat{\boldsymbol f}_L$ is the time Laplace transform of the lift force:
\begin{equation}
\boldsymbol f_L(\boldsymbol\xi(t),t)=\frac{6.46\,\beta}{ 2\,\pi r_p}\sqrt{\frac{\nu}{||\boldsymbol\omega(\boldsymbol\xi(t),t)||}} [\boldsymbol v(t)-\boldsymbol u(\boldsymbol\xi(t),t)]\times\boldsymbol\omega(\boldsymbol\xi(t),t)
\end{equation}
A straightforward application of the generalized Taylor's formula (\ref{GenTaylor}) yields:
 \begin{eqnarray}
 \label{MRSaffmanTaylor}      \D=D_0\mathds     1&+&\operatorname{Sym}\int_{t_0}^\infty      \mathrm{d}t\,
\left\langle\left[\delta \boldsymbol{u}(\boldsymbol{\xi}(t),t)
+\beta\tau\,\delta\frac{\mathrm{D}\boldsymbol{u}}{\mathrm{D}t}(\boldsymbol{\xi}(t),t) 
+\frac{1}{6}  r_p^2 \, \delta\nabla^2 \boldsymbol{u}(\boldsymbol{\xi}(t),t)+\tau\delta \boldsymbol f_L(\boldsymbol\xi(t),t)\right]\right.
\nonumber\\
&\otimes&\left.\left[\delta \boldsymbol{u}(\boldsymbol{\xi}(t_0),t_0)
+\beta\tau\,\delta\frac{\mathrm{D}\boldsymbol{u}}{\mathrm{D}t}(\boldsymbol{\xi}(t_0),t_0)    
+\frac{1}{6}    r_p^2  \,\,\delta  \nabla^2 \boldsymbol{u}(\boldsymbol{\xi}(t_0),t_0)+\tau\delta \boldsymbol f_L(\boldsymbol\xi(t_0),t_0)
\right]\right\rangle\nonumber\\
 \end{eqnarray} 
 
Eq. (\ref{MRSaffmanTaylor}) allows evaluating how the autocorrelation of the lift force and its cross-correlations with the other terms contribute to the eddy diffusivity. This can be carried out by the analysis of  trajectories from available RADAR data or numerical simulations.

It should be noted that Eqs. (\ref{MRSaffman}) and (\ref{MRSaffmanTaylor})  do not contain lift terms depending on the angular velocity $\boldsymbol\Omega_p$ of the particle, the so-called Magnus effect. Indeed, among higher order corrections (see. Eqs. (2.17)-(4.15) in \cite{Saffman} and Eq. (4) in \cite{Lift}), a lift force acting on the particle of the form \cite{Magnus}:
\begin{equation}
\label{Magnus}
\rho_f r_p^3\pi\boldsymbol\Omega_p\times[\boldsymbol v(t)-\boldsymbol u(\boldsymbol\xi(t),t)]
\end{equation}
should be added to Eq. (\ref{LiftForce}). However, the ratio between this term per unity of mass and the Stokes drag is:
\begin{equation}
\label{Magnus}
\frac{\frac{2}{3}\beta \pi r^3_p||\boldsymbol\Omega_p\times[\boldsymbol v(t)-\boldsymbol u(\boldsymbol\xi(t),t)]||}{\frac{4}{3}\pi r_p^3 ||\boldsymbol v(t)-\boldsymbol u(\boldsymbol\xi(t),t) ||/\tau}\leq\beta||\boldsymbol\Omega_p||\tau\leq\beta\frac{||\boldsymbol\Omega_p||}{||\boldsymbol\omega||}\text{St}
\end{equation}
This ratio turns out to be of order $O$(St), while the one between Saffman lift and Stokes drag was $\sim O(\sqrt{\text St})$. This justifies why the Magnus term (\ref{Magnus}) is often neglected for small solid particles, unless the angular velocity is  high. However, for a freely rotating sphere, $\boldsymbol\Omega_p=1/2\,\boldsymbol\omega$ \cite{Saffman}. We did not take that term into account here for the sake of simplicity, it being often ignored. In any case, its inclusion in Eq. (\ref{MRSaffmanTaylor}) is trivially inside the addend $\boldsymbol f_L$.  
}

\section{Conclusions} 
\label{sec:end}

We analyzed general conditions under which a {\color{black} generalized}  Taylor's eddy diffusivity 
formula applies to inertial particle models. 

It is worth emphasizing that Taylor's formula for the Basset--Boussinesq--Oseen
model of inertial particle dynamics with the inclusion of the Brownian force, 
is formally the same  as the Taylor's formula for  tracer particles. The equivalence  is, however,  
only  formal. Since  the  time integral of the fluid velocity autocorrelation function is carried out
along particle trajectories, the well-known mismatch between fluid and
particle   trajectories   leads   in   general   to   different   eddy
diffusivities.

In the case of the Maxey-Riley model, new terms appear in the expression for
the eddy  diffusivity with respect to  the tracer case and  thus with
respect to  the Basset--Boussinesq--Oseen model.  We also discussed under which coditions 
the two models admit the  same formal expression  for the eddy diffusivity. {\color{black} Similar conclusions were drawn taking into account lift forces.}

Our analysis encompasses,  as special  cases of
interest  in  applications,  the  two relevant  examples  of  particle
dynamics  forced  by the  Coriolis  contribution  (for application  to
dispersions  in   geophysical  flows)  and  the   Lorentz  force  (for
application  to  dispersions  of  charged  particles  in  electrically
neutral flows).  In this latter case,  we proved that in  the limit of
small inertia (i.e. $\mathrm{St}\,\downarrow\, 0$) and magnetic field $\boldsymbol B^{*}$ such that  $\|\boldsymbol{B}^{*}\|$ is independent
of $\mathrm{St}$,  the inertial  particle dynamics reduces  to a  tracer dynamics
with  a  carrier  flow  which  now  becomes  compressible.  Clustering
phenomena  induced  by  the  magnetic   field  are  thus  expected  to
emerge. For a  vanishing carrier flow, the combined  roles of Brownian
motion and  magnetic field has been  proved to give rise  to a smaller
eddy  diffusivity  than  the molecular  diffusivity  $D_0$.  Transport
depletion  is thus  expected  in applications  involving the  magnetic
field.  Similar   conclusions  can   be  obtained  for   the  Coriolis
contribution. The mathematical  structure of this term  is indeed very
similar to the Lorentz force.
Taylor's formula for  tracer dispersion has ubiquitous applications in
the study of turbulent transport.  We thus expect  that our analysis  will
be useful for further investigations  of large-scale  transport properties of 
inertial  particles  under  the action  of different forcing mechanisms.

\acknowledgments
The research of S. Boi is funded by the AtMath Collaboration at the University of Helsinki. 
P. Muratore-Ginanneschi also acknowledges support from the AtMath Collaboration.

\appendix*

\section*{Appendix A}
\label{app:proof}
\setcounter{equation}{0}

\begin{proposition}
Under the hypotheses~\textbf{I-II-III}, {\color{black} generalized}  Taylors's identity (\ref{GenTaylor})
holds true for any dynamical model of the form (\ref{GT:eq})
\end{proposition}
To prove the claim, we need first to couch (\ref{GT:def}) into the form (\ref{GT:adapted})
which is more adapted to discuss the large time limit. 
This is done by first applying to (\ref{GT:def}) the double integral inversion formula over 
a triangular domain
\begin{eqnarray}
\lefteqn{
\operatorname{Sym} \int_0^t      \mathrm{d}s\,     
\langle\,\delta\boldsymbol{v}(t)\otimes\delta\boldsymbol{v}(s)\rangle
=}
\nonumber\\
&&\sum_{i j=0}^{N}\operatorname{Sym}\int_{0}^{t}\mathrm{d}s_{3}\,
\int_{0}^{t}\mathrm{d}s_{2}\int_{s_{2}}^{t}\mathrm{d}s_{1}\,\mathds{K}_{i}(t-s_{3})\,
\tilde{\mathds{C}}_{i j}(s_{3},s_{2})\,\mathds{K}_{j}^{T}(s_{1}-s_{2})
\label{app:def}
\end{eqnarray}
Performing the sequence the change of variables $s_{1}=u_{1}+s_{2}$, $s_{2}=t-u_{2}$ and
$s_{3}=t-u_{3}$ yields (\ref{GT:adapted}) which, for reading convenience, we re-write 
here as
\begin{eqnarray}
\lefteqn{
\operatorname{Sym} \int_0^t      \mathrm{d}s\,     
\langle\,\delta\boldsymbol{v}(t)\otimes\delta\boldsymbol{v}(s)\rangle
=}
\nonumber\\
&&\sum_{i j=0}^{N}\operatorname{Sym}\int_{0}^{t}\mathrm{d}u_{3}\,
\int_{0}^{t}\mathrm{d}u_{2}\int_{0}^{u_{2}}\mathrm{d}u_{1}\,\mathds{K}_{i}(u_{3})\,
\tilde{\mathds{C}}_{i j}(t-u_{3},t-u_{2})\,\mathds{K}_{j}^{T}(u_{1})
\label{app:adapted}
\end{eqnarray}
We now invoke hypotheses~\textbf{I-II}. They ensure
that (\ref{app:adapted}) (or equivalently (\ref{GT:adapted})) is absolutely integrable in the large time limit. 
Before proving this claim, it is convenient to proceed to analyze its implications. 
Namely, if we take the limit under the integral and invoke
hypothesis~\textbf{III}, upon applying once again the double integral inversion formula over a triangular domain, 
we obtain
\begin{eqnarray}
\label{}
\lefteqn{
\mathds{D}=\lim_{t\uparrow\infty}\operatorname{Sym} \int_0^t      \mathrm{d}s\,     
\langle\,\delta\boldsymbol{v}(t)\otimes\delta\boldsymbol{v}(s)\rangle
}
\nonumber\\&&
=\sum_{i j=0}^{N}\operatorname{Sym}\int_{0}^{\infty}\mathrm{d}u_{3}\,
\int_{0}^{\infty}\mathrm{d}u_{1}\,\mathds{K}_{i}(u_{3})\,
\mathds{F}_{i j}(u_{3},u_{1})\,\mathds{K}_{j}^{T}(u_{1})
\label{app:limit}
\end{eqnarray}
where by (\ref{GT:property})
\begin{eqnarray}
\label{}
\mathds{F}_{i j}(u_{3},u_{1})=\int_{u_{1}}^{\infty}\mathrm{d}u_{2}
\left\{
\begin{array}{ll}
\mathds{C}_{i j}(u_{2}-u_{3}) \hspace{0.2cm}&\hspace{0.2cm} \forall\, u_{2}\geq u_{3}
\\[0.3cm]
\mathds{C}_{j i}^{T}(u_{3}-u_{2}) \hspace{0.2cm}&\hspace{0.2cm}\forall\,  u_{2}< u_{3}
\end{array}
\right.
\label{app:kernel}
\end{eqnarray}
The kernel (\ref{app:kernel}) is in fact a function of $u_{1}-u_{3}$ alone and admits important simplifications. 
Namely, we notice that for $u_{1}\,\geq\,u_{3}$
\begin{eqnarray}
\mathds{F}_{i j}(u_{3},u_{1})
=\int_{0}^{\infty}\mathrm{d}u\,\mathds{C}_{i j}(u)-\int_{0}^{u_{1}-u_{3}}\mathrm{d}u\,\mathds{C}_{i j}(u)
\nonumber
\end{eqnarray}
whilst for $u_{3}\,>\,u_{1}$ we find
\begin{eqnarray}
\mathds{F}_{i j}(u_{3},u_{1})=\int_{0}^{\infty}\mathrm{d}u\,
\mathds{C}_{i j}(u)+\int_{0}^{u_{3}-u_{1}}\mathrm{d}u\,\mathds{C}_{j i}^{T}(u)
\nonumber
\end{eqnarray}
Upon gleaning these observations, we conclude after a further application of (\ref{GT:property}) 
that
\begin{eqnarray}
\label{app:newkernel}
\mathds{F}_{i j}(u_{3},u_{1})=\int_{0}^{\infty}\mathrm{d}u\,\mathds{C}_{ i j}(u)
-\int_{0}^{u_{1}-u_{3}}\mathrm{d}u\,\mathds{C}_{i j}(u)\equiv\mathds{\hat{C}}_{i j}(0)
-\tilde{\mathds{F}}_{i j}(u_{1}-u_{3})
\end{eqnarray}
where furthermore
\begin{eqnarray}
\tilde{\mathds{F}}_{i j}(-t)=\int_{0}^{-t}\mathrm{d}u\,
\mathds{C}_{i j}(u)=-\int_{0}^{t}\mathrm{d}u\,
\mathds{C}_{i j}(-u)=-\int_{0}^{t}\mathrm{d}u\,
\mathds{C}_{j i}^{T}(u)=-\tilde{\mathds{F}}_{j i}^{T}(t)
\label{app:property}
\end{eqnarray}
We have now forged all the tools needed to prove the proposition. 
If we take the $\operatorname{Sym}$ operation under the integral sign
and rename {\color{black}dummy integration/summation variables}, we get into
\begin{eqnarray}
\mathds{D}
=\sum_{i j=0}^{N}\int_{0}^{\infty}\mathrm{d}u_{3}
\int_{0}^{\infty}\mathrm{d}u_{1}\,\mathds{K}_{i}(u_{3})\, \frac{\mathds{F}_{i j}(u_{3},u_{1})+\mathds{F}_{j i}^{T}(u_{1},u_{3})}{2}
\,\mathds{K}_{j}^{T}(u_{1})
\label{app:final}
\end{eqnarray}
In view of (\ref{app:newkernel}), (\ref{app:property}) the chain of identities
\begin{eqnarray}
\lefteqn{
\frac{\mathds{F}_{i j}(u_{3},u_{1})+\mathds{F}_{j i}^{T}(u_{1},u_{3})}{2}=
}
\nonumber\\&&
\frac{\mathds{\hat{C}}_{i j}(0)+\mathds{\hat{C}}_{j i}^{\mathsf{T}}(0)}{2}
-
\frac{\mathds{\tilde{F}}_{i j}(u_{1}-u_{3})+\mathds{\tilde{F}}_{j i}^{T}(u_{3}-u_{1})}{2}=
\nonumber\\&&
\frac{\mathds{\hat{C}}_{i j}(0)+\mathds{\hat{C}}_{j i}^{\mathsf{T}}(0)}{2}
-
\frac{\mathds{\tilde{F}}_{i j}(u_{1}-u_{3})-\mathds{\tilde{F}}_{i j}(u_{1}-u_{3})}{2}
=\frac{\mathds{\hat{C}}_{i j}(0)+\mathds{\hat{C}}_{j i}^{\mathsf{T}}(0)}{2}
\label{app:final}
\end{eqnarray}
holds true. Hence, the kernel in (\ref{app:final}) is independent of the integration variables $u_{1}$, $u_{3}$ and 
the double integral factorizes in the product of two integrals. As a consequence, (\ref{app:final}) reduces to {\color{black} generalized} Taylor's 
formula (\ref{GenTaylor}), as claimed.

Finally, we can return to the proof that hypotheses~\textbf{I-II} are sufficient to guarantee that
it is safe to apply the dominated convergence theorem to (\ref{app:adapted}).
Namely, the chain of inequalities
\begin{eqnarray}         
\lefteqn{\left|
\int_0^t          \mathrm{d}s
\Big{\langle} \delta{v}^{m}(t)\;\delta{v}^{n}(s)
\Big{\rangle}\,\right|
}
\nonumber\\  
&& \leq \sum_{i j=0}^{N}
\int_0^t  \mathrm{d}s_{3} \int_0^{t}  \mathrm{d}s_1  \int_0^{s_{1}} \mathrm{d}s_2 \, 
|\mathds{K}_{i}^{m l}(s_{3})|\,|\mathds{K}_{j}^{n k}(s_{1})|\,   |\tilde{\C}^{l\,k}_{i\,j}(s_{3},s_{2})|
\nonumber\\
&&\leq
\int_0^t  \mathrm{d}s_{3} \int_0^{t}  \mathrm{d}s_1   V^{m}(s_{3})\,V^{n}(s_{1})\,\int_0^{s_{1}} \mathrm{d}s_2 \,F (s_{3}-s_{2})
\label{app:bound}
\end{eqnarray}
holds for
\begin{eqnarray}
V^{n}(t)=\sum_{i=0}^{N}\sum_{l=1}^{d}|\mathds{K}_{i}^{n l}(t)|
\nonumber
\end{eqnarray}
The innermost integral in (\ref{app:bound}) satisfies
\begin{eqnarray}
0\leq \int_{0}^{s_{1}} \mathrm{d}s_2 \,F (s_{3}-s_{2})\,<\, \int_{-\infty}^{\infty} \mathrm{d}s \,F (s)\equiv 2\,f_{\star}
\nonumber
\end{eqnarray}
since $F$ is positive, even and integrable by hypothesis. We conclude that
{\color{black}
\begin{eqnarray}
\lim_{t\uparrow\infty}\left|
\int_0^t          \mathrm{d}s
\Big{\langle} \delta{v}^{m}(t)\;\delta{v}^{n}(s)
\Big{\rangle}\,\right|\,<\, 2\, [(N+1)\,K_{\star}\, d]^{2}\,f_{\star}\,<\,\infty\, ,
\nonumber
\end{eqnarray}
with $K_*$ being defined in Eq. (\ref{KappaStar}).
}

\newpage
{\color{black}
\begin{turnpage}
\begin{table}[h]
\begin{tabular}{|c|c|} 
\multicolumn{2}{c}{}\\
\multicolumn{2}{c}{\textbf{ TABLE OF GENERALIZED TAYLOR FORMULAE}}\\
\multicolumn{2}{c}{}\\
\hline
Model                          & Eddy diffusivity  \\ 
\hline
\hline
Basset-Boussinesq-Oseen equation &  $\lim_{t\uparrow\infty }\operatorname{Sym} \int_0^t      \mathrm{d}s\,     
 \langle\,\delta\boldsymbol{u}(\boldsymbol \xi(s),s) \otimes\delta\boldsymbol{u}(\boldsymbol \xi(t),t) \rangle \;\;\;$\\
 plus white noise and constant gravity & \\
\hline
Basset-Boussinesq-Oseen equation &  $\frac{D_0}{\tau^2} \mathds{\hat{A}}^{-1}(0)\,
(\mathds{\hat{A}}^{-1})^{T}(0)+  \frac{1}{\tau^2}  \mathds{\hat{A}}^{-1}(0) 
\operatorname{Sym}\Big{(}\hat{\C}_{0 0} (0)\Big{)} \, (\mathds{\hat{A}}^{-1})^{T}(0)$\\
 plus white noise and Lorentz force & where \\
 &$ (\mathds{\hat{A}}^{-1})^{\mu\nu}(z)=\frac{1}{a^{2}(z)+\gamma^{2}\,\|\boldsymbol{B}\|^{2}}
\left[a(z)\,\delta^{\mu\nu}-\gamma\,B^{i}\,\epsilon^{\mu \sigma \nu}+\frac{\gamma^{2}}{a(z)}B^{k}\,B^{j}\right]$\\ 
 & $a(z)=z+\frac{1}{\tau}+ \sqrt{\frac{3\beta z}{\pi\tau}}$ \\ & $\gamma=\frac{q}{\frac{4}{3}\pi r^3_p\rho_p}$ \\
 &  \\
\hline
Basset-Boussinesq-Oseen equation &  \\
plus white noise and Lorentz force & $\text{diag}\left(\frac{D_0}{1+\gamma^{2}\,B^2\,\tau^2},\frac{D_0}{1+
\gamma^{2}\, B^2 \,\tau^2},D_0\right)$ \\
 (flow at rest, constant magnetic field $\boldsymbol B=(0,0,B)$ ) & \\
 \hline
Basset-Boussinesq-Oseen equation & $D_0\; \mathds{A}^{-1} ({\mathds A}^{-1})^{T}+
\mathds{A}^{-1}   \operatorname{Sym}\Big{(}\hat{\C}_{0 0}(0)\Big{)}   ({\mathds
A}^{-1})^{T}$ \\
plus white noise and Lorentz force &  where \\
(limit of vanishing Stokes number and $\boldsymbol B*=\gamma\tau\boldsymbol B$ ) & $\mathds{A}^{\mu\nu}=\delta^{\mu\nu}+B^{*\,
i}\,\epsilon^{\mu \sigma \nu}$ \\
 \hline
  \end{tabular} 
\end{table}
\end{turnpage}

\begin{turnpage}
\begin{table}[h]
\begin{tabular}{|c|c|} 
\hline
Model                          & Eddy diffusivity  \\ 
\hline
\hline
Basset-Boussinesq-Oseen equation & $\frac{D_0}{\tau^2}
 \mathds{\hat{A}}^{-1}(0)\,({\mathds{\hat{A}}}^{-1})^{T}(0)+ 
\mathds{\hat{A}}^{-1}(0)\,\mathds{\hat{B}}(0)\, 
\operatorname{Sym}\Big{(}\hat{\C}_{0 0}(0)\Big{)}   \,\mathds{\hat{B}}^T(0)\, ({\mathds{\hat{A}}}^{-1})^{T}(0)$ \\
plus constant gravity, white noise and Coriolis force, &  where \\
Basset force neglected & $ \mathds{\hat{A}}^{\mu\nu}(z)=
\left(z+\frac{1}{\tau}\right)\,\delta^{\mu\nu}
+2\,\Omega^{\sigma}\,      \epsilon^{\mu \sigma \nu}$ \\
& $\mathds{\hat{B}}^{\mu\nu}(z)=
\left(\beta\,z+\frac{1}{\tau}\right)\,\delta^{\mu\nu}
+2\,\beta\,\Omega^{\sigma}\, \epsilon^{\mu \sigma \nu}$ \\
 \hline
 Basset-Boussinesq-Oseen equation &  \\
plus white noise, constant gravity and Coriolis force & $\text{diag}\left(\frac{D_0}{1+4\,\Omega^2\,\tau^2},\frac{D_0}{1+
4\, \Omega^2 \,\tau^2},D_0\right)$ \\
 (flow at rest,constant angular velocity $\boldsymbol \Omega=(0,0,\Omega)$ ) & \\
 \hline
 Basset-Boussinesq-Oseen equation & $\frac{D_0}{1+4/\Ro^2}\mathds    1+   \boldsymbol
M\otimes\boldsymbol M\int_0^\infty\mathrm{d} t\, C(t)$ \\
plus white noise, constant gravity and Coriolis force, & where \\
Basset force neglected. & $\boldsymbol{M}= \frac{1}{1+4/\Ro^2}\left(\begin{aligned}  &  1+4\beta/\Ro^2  \\  &
2(\beta-1)/\Ro \\ & 0 \\
\end{aligned}\right)$ \\
 (shear flow $\boldsymbol u(\boldsymbol{x}, t)= u(x_{2}, x_{3},t) \boldsymbol
e_{1}\; ,$  &  $C(t)=\lim_{t^\prime\to\infty}\langle\delta u(\boldsymbol\xi(t^\prime),t^\prime)\delta u(\boldsymbol\xi(t+t^\prime),t+t^\prime) \rangle$ \\
 constant angular velocity $\boldsymbol \Omega=(0,0,\Omega)$ , & \\
 fixed Rossby number $\mathrm{Ro}=1/(\tau\,\Omega)$ ) & \\
 \hline
  \end{tabular} 
\end{table}
\end{turnpage}

\begin{turnpage}
\begin{table}[h]
\vspace*{-5cm}\begin{tabular}{|c|c|} 
\hline
Model                          & Eddy diffusivity  \\ 
\hline
\hline
 Maxey-Riley equation & $D_0\mathds     1+\operatorname{Sym}\int_{t_0}^\infty      \mathrm{d}t\,$
\\
 (including white noise, Fax\'en, and Auton terms) & $\left\langle\left[\delta \boldsymbol{u}(\boldsymbol{\xi}(t),t)
+\beta\tau\,\delta\frac{\mathrm{D}\boldsymbol{u}}{\mathrm{D}t}(\boldsymbol{\xi}(t),t) 
+\frac{1}{6}  r_p^2 \, \delta\nabla^2 \boldsymbol{u}(\boldsymbol{\xi}(t),t)\right]\right. $ \\
 & $\otimes\left.\left[\delta \boldsymbol{u}(\boldsymbol{\xi}(t_0),t_0)
+\beta\tau\,\delta\frac{\mathrm{D}\boldsymbol{u}}{\mathrm{D}t}(\boldsymbol{\xi}(t_0),t_0)    
+\frac{1}{6}    r_p^2  \,\,\delta  \nabla^2 \boldsymbol{u}(\boldsymbol{\xi}(t_0),t_0)
\right]\right\rangle$ \\
 \hline
 Maxey-Riley equation + lift force & $ \D=D_0\mathds     1+\operatorname{Sym}\int_{t_0}^\infty      \mathrm{d}t\,$ \\
 & $\left\langle\left[\delta \boldsymbol{u}(\boldsymbol{\xi}(t),t)
+\beta\tau\,\delta\frac{\mathrm{D}\boldsymbol{u}}{\mathrm{D}t}(\boldsymbol{\xi}(t),t) \right.\right.$\\
&$\left.\left.+\frac{1}{6}  r_p^2 \, \delta\nabla^2 \boldsymbol{u}(\boldsymbol{\xi}(t),t)+\tau\delta \boldsymbol f_L(\boldsymbol\xi(t),t)\right]\right.$\\
& $\otimes\left.\left[\delta \boldsymbol{u}(\boldsymbol{\xi}(t_0),t_0)
+\beta\tau\,\delta\frac{\mathrm{D}\boldsymbol{u}}{\mathrm{D}t}(\boldsymbol{\xi}(t_0),t_0)    \right.\right.$\\
&$\left.\left.+\frac{1}{6}    r_p^2  \,\,\delta  \nabla^2 \boldsymbol{u}(\boldsymbol{\xi}(t_0),t_0)+\tau\delta \boldsymbol f_L(\boldsymbol\xi(t_0),t_0)
\right]\right\rangle$ \\
\hline
\end{tabular} 
\end{table}
\end{turnpage}
}

\FloatBarrier

\end{document}